\documentstyle[11pt,amssymb,newlfont]{amsart}
\newcommand{\R}{\Bbb R}
\newcommand{\N}{\Bbb N}
\newcommand{\diver}{\operatorname{div}}
\newcommand{\0}{\emptyset}

\newcommand{\be}[1]{\begin{equation}\label{#1}}
\renewcommand{\phi}{\varphi}
\newcommand{\eps}{\varepsilon}

\newcommand{\bml}[1]{\begin{multline}\label{#1}}
\newcommand{\bes}{\begin{equation*}}
\newcommand{\bs}{\begin{split}}
\newcommand{\bdm}{\begin{displaymath}}
\newcommand{\edm}{\end{displaymath}}

\newcommand{\di}{\partial}
\renewcommand{\phi}{\varphi}
\newtheorem{th}{Theorem}[section]
\newtheorem{lem}{Lemma}[section]

\begin{document}
\title[]{Coupled problems on stationary flow of electrorheological fluids 
under the conditions of nonhomogeneous temperature distribution.}

\author[]{R.H.W. Hoppe, W.G. Litvinov.}
\date{}
\address{\newline
Lehrstuhl f\"ur Angewandte Analysis mit Schwerpunkt Numerik
\hspace*{3ex}\newline
Universit\"at Augsburg
\newline Universit\"atsstrasse, 14
\newline 86159 Augsburg, Germany
\hspace*{3ex}\newline
E-mails:\newline
hoppe math.uni-augsburg.de\newline
litvinov math.uni-augsburg.de.} 

\begin{abstract}

We set up and study a coupled problem on stationary non-isothermal flow of
electrorheological fluids. The problem consist in finding functions of 
velocity, pressure and temperature which satisfy the motion equations, the
condition of incompressibility, the equation of the balance of thermal 
energy and boundary conditions.

We introduce the notions of a $P$-generalized solution and generalized 
solution of the coupled problem. In case of the $P$-generalized solution the 
dissipation of energy is defined by the regularized velocity field, which
leads to a nonlocal model.

Under weak conditions,  we prove the existence of the $P$ -generalized 
solution of the coupled problem. The existence of the generalized solution is 
proved under the conditions on smoothness of the boundary and on smallness
of the data of the problem

\end{abstract}
\maketitle

\makeatletter\@addtoreset{equation}{section}\makeatother
\def\theequation{\arabic{section}.\arabic{equation}}

\section{Introduction}
Electrorheological fluids are smart materials which
are concentrated suspensions of polarizable particles 
in a nonconducting dielectric liquid. In moderately large electric fields, the
particles form chains along the field lines, and these chains then aggregate to 
form columns \cite{11}. These chainlike and columnar 
structures cause dramatic changes in
the rheological properties of the suspensions. The fluids become anisotropic,
the apparent viscosity (the resistance to flow) in the direction orthogonal to
the direction of electric field abruptly increases, while the apparent viscosity
in the direction of the electric field changes not so drastically.

The chainlike and columnar structures are destroyed under the action of large 
stresses, and then the apparent viscosity of the fluid decreases and the fluid
becomes less anisotropic.

On the basis of experimental results the following constitutive equation
was developed in \cite{5}:
\begin{equation}\label{1.1}
\sigma_{ij}(p,u,E)=-p\delta_{ij}+2\phi(I(u),\vert E\vert,\mu(u,E))\eps_{ij}
(u), \quad i,j=1,\dots,n, \quad n=2 \mbox{ or } 3.
\end{equation}
Here, $\sigma_{ij}(p,u,E)$ are the components of the stress tensor which 
depend on the pressure $p$, the velocity vector $u=(u_1,\dots,u_n)$ and the 
electric field strength $E=(E_1,\dots,E_n)$, $\delta_{ij}$ are the components
of the unit tensor (the Kronecker delta), and 
$\eps_{ij}(u)$ are the components of the rate of strain tensor
\begin{equation}\label{1.2}
\eps_{ij}(u)=\frac12\big(\frac{\di u_i}{\di x_j}\,+\,\frac{\di u_j}{\di x_i}
\big).
\end{equation}
Moreover, $I(u)$ is the second invariant of the rate of strain tensor
\begin{equation}\label{1.3}
I(u)=\sum_{i,j=1}^n (\eps_{ij}(u))^2,
\end{equation}
and $\phi$ the viscosity function depending on $I(u)$, $\vert E\vert$ and $\mu(
u,E)$, where
\begin{equation}\label{1.4}
(\mu(u,E))(x)=\Big(\frac{u(x)}{\vert u(x)\vert},\,\,\frac{E(x)}{\vert E(x)
\vert}\Big)_{\R^n}^2\,=\,\frac{(\sum_{i=1}^n u_i(x)E_i(x))^2}{(\sum_{i=1}^n
(u_i(x))^2)(\sum_{i=1}^n(E_i(x))^2)}.
\end{equation}
So $\mu(u,E)$ is the square of the scalar product of the unit vectors $\frac
{u}{\vert u\vert}$ and $\frac E{\vert E\vert}$. The function $\mu$ is defined
by \eqref{1.4} in the case of an immovable frame of reference. If the frame
of reference moves uniformly with a constant velocity $\check u=(\check u_1,
\dots,\check u_n)$, then we set:
\begin{equation}\label{1.5}
\mu(u,E)(x)=\Big(\frac{u(x)+\check u}{\vert u(x)+\check u\vert},\,\,\frac
{E(x)}{\vert E(x)\vert}\Big)_{\R^n}^2.
\end{equation}
As the scalar product of two vectors is independent of the frame of reference,
the constitutive equation \eqref{1.1} is invariant with respect to the group
of Galilei transformations of the frame of reference that are represented as
a product of time-independent translations, rotations and uniform motions.

It is obvious that $\mu(u,E)(x)\in[0,1]$, and for fixed
$y_1,y_2\in\R_+$, where $\R_+=\{z\in\R,\,\,\,z\ge 0\}$, the function $y_3\to
\phi(y_1,y_2,y_3)$ reaches its maximum at $y_3=0$ and its minimum at $y_3=1$ 
when the vectors $u(x)+\check u$ and $E$ are  correspondingly orthogonal 
and parallel.

The function $\mu$ defined by \eqref{1.4}, \eqref{1.5} is not specified at
$E=0$ and at $u=0$, and there does not exist an extension of $\mu$
by continuity 
to the values of $u=0$ and $E=0$. However, at $E=0$ there is no influence of 
the electric field. Therefore,
\begin{equation}\label{1.6}
\phi(y_1,0,y_3)=\tilde\phi(y_1),\qquad  y_3\in[0,1],
\end{equation}
and the function $\mu(u,E)$ need not be specified at $E=0$. Likewise, in case 
that the 
measure of the set of points $x$ at which $u(x)=0$ is zero, the function $\mu$ 
need not also be specified at $u=0$. But in the general the  function $\mu$
is defined as follows:
\begin{equation}\label{1.7}
\mu(u,E)(x)=\Big(\frac{\alpha\tilde I+u(x)+\check u}{\alpha\sqrt n+\vert u(x)+
\check u\vert},\,\,\frac{E(x)}{\vert E(x)\vert}\Big)_{\R^n}^2,
\end{equation}
where $\tilde I$ denotes  a vector with components equal to one, and $\alpha$ 
is a small
positive constant. If $u(x)\ne 0$ almost everywhere in $\Omega$, we may choose
$\alpha=0$.

The viscosity function $\phi$ is identified by approximation of flow curves, 
see \cite{5}, and it was shown in \cite{5} that it can be represented as 
follows:

\begin{equation}\label{1.8}
\phi(I(u),\vert E\vert,\mu(u,E))=b(\vert E\vert,\mu(u,E))(\lambda+I(u))^
{-\frac12}+\psi(I(u),\vert E\vert,\mu(u,E)),
\end{equation}
where $\lambda$ is a small parameter, $\lambda\ge 0$.

In the special case that the direction of the velocity vector $u(x)$ at each
point $x$ at which $E(x)\ne 0$ is known, the function $x\to\mu(u,E)(x)$ 
becomes well-known, and the viscosity functions \eqref{1.8} takes the form
\begin{equation}\label{1.9}
\phi(I(u),\vert E\vert,x)=e(\vert E\vert,x)(\lambda+I(u))^{-\frac12}+
\psi_1(I(u),\vert E\vert,x),
\end{equation} 
where
\begin{align}
&e(\vert E\vert,x)=b(\vert E\vert,\mu(u,E)(x)), \notag \\
&\psi_1(I(u),\vert E\vert,x)=\psi(I(u),\vert E\vert,\mu(u,E)(x)). \label{1.10}
\end{align}
Here and thereafter the function of electric field $ E$ is assumed to be known.
The equations for the functions $E$ and $(p,v)$ are separated, see \cite{5}.

In actual practice, the temperature fields in electrorheological fluids are 
nonuniform, and in many cases this non-homogeneity drastically affects on the 
flow of electrorheological fluids. Because of this, we reckon below that the
functions $b,\psi,e,\psi_1$ in \eqref{1.8} and \eqref{1.9} depend also on the
temperature $\tau$. Therefore, we consider the following viscosity functions
\begin{align}
&\phi_1=b(\vert E\vert,\mu(u,E),\tau)(\lambda+I(u))^{-\frac12}+\psi(I(u),
        \vert E\vert,\mu(u,E),\tau), \label{1.11}\\
&\phi_2=e(\vert E\vert,\tau,x)(\lambda+I(u))^{-\frac12}+\psi_1(I(u),\vert E
\vert,\tau,x),  \label{1.12}
\end{align}
and the stress tensor is defined by
\begin{equation}\label{1.13}
\sigma_{ij}=-p\delta_{ij}+2\phi_k\eps_{ij}(u), \quad k=1 \text{ or } 2,
\quad i,j=1,\dots,n.
\end{equation}
Further we study coupled problems on non isothermal flow of electrorheological fluids with the constitutive equation \eqref{1.13} for $k$ equals one and two.  

In Sections 2 and 3 we introduce equations governing the process of 
non isothermal flows of electrorheological fluids. The coupled problem on 
non-isothermal flow consists in finding functions of velocity, pressure and
temperature which satisfy the motion equations, the condition of
incompressibility, the equation of the balance of thermal energy and boundary
conditions.

We introduce the notions of a $P$-generalized and generalized solution of the 
coupled problem. In case of the $P$-generalized solution the dissipation of 
energy is defined by the regularized velocity field, which leads to a nonlocal model.

Some auxiliary results are set in Section 4, and the existence of a 
$P$-generalized solution of the coupled problem for the viscosity function
$\phi_2$ is proved in Section 5.

Assuming that the data of the problem are small and the boundary of the domain
under consideration is smooth, we prove in Section 6 the existence of a
generalized solution of the coupled problem for the viscosity function 
$\phi_2$.

In Section 7 we consider coupled problem for the viscosity function 
$\phi_1$ and formulate a result on the existence of the P-generalized 
solution for $\phi_1$. A proof of this result is contained in Sections 
8 and 9.

\section{Equations of flow and thermal balance, boundary conditions.}

We consider stationary problem on non-isothermal flow of the electrorheological
fluid. The inertia forces are assumed to be small as compared with the 
internal forces caused by the viscous stresses. Then the equations of motion 
take the following form:
\begin{equation}\label{2.1}
\frac{\di p}{\di x_i}-2\,\frac{\di}{\di x_j}[\phi_k\eps_{ij}(u)]=K_i \quad
             \text{in } \Omega, \quad i=1,\dots,n, \quad k=1 \text{ or }2,
\end{equation}
where $K_i$ are the components of the volume force vector $K$. In \eqref{2.1}
and below the Einstein convention on summation over repeated index is applied.

The velocity function $u$ meets the incompressibility condition
\begin{equation}\label{2.2}
\diver u=\sum_{i=1}^n \frac{\di u_i}{\di x_i}=0 \quad \text{in }\Omega.
\end{equation}
The equation of the balance of thermal energy is the following:
\begin{equation}\label{2.3}
\chi\sum_{i=1}^n \frac{\di^2\tau}{\di x_i^2}+2\eps\phi_k I(u)-u_i\frac
{\di\tau}{\di x_i}=0 \quad\text{in } \Omega, \quad k=1\text{ or } 2.
\end{equation}
Here $\chi$ is the thermal diffusivity, $\eps=(\rho c_p)^{-1}$, where $\rho$
is the density, $c_p$ specific heat.
We reckon that $\chi$  and $\eps$ are positive constants.

The first, second and third terms in the left-hand side of \eqref{2.3}
determine the increments of the temperature in the unit of time produced
respectively by heat conduction, dissipation of energy and thermal convection.

We assume that $\Omega$ is a bounded domain in $\R^n$, $n=2$ or 3 with a
Lipschitz continuous boundary $S$. Suppose that $S_1$ and $S_2$ are open subsets
of $S$ such that $S_1$ is non-empty, $S_1\cap S_2=\0$, and $\overline 
S_1\cup\overline S_2=S$. We consider mixed boundary conditions for the 
functions $u,p$, wherein velocities are specified on $S_1$  and surface forces 
are given on $S_2$, i.e.
\begin{gather}
u\Big\vert_{S_1}=\hat u, \label{2.4}\\
[-p\delta_{ij}+\phi_k\eps_{ij}(u)]\nu_j\Big\vert_{S_2}=F_i, \quad i=1,\dots,n, 
             \quad  k=1 \text{ or } 2. \label{2.5}
\end{gather}
Here $F_i$ and $\nu_j$ are components of the vectors of surface force $F=
(F_1,\dots,F_n)$ and the unit outward normal $\nu=(\nu_1,\dots,\nu_n)$ to $S$, 
respectively.

Temperature field on the boundary $S$ is considered to be given
\begin{equation}\label{2.6}
\tau\Big\vert_{S}=\hat\tau.
\end{equation}
We assume that $\hat u\in H^{\frac12}(S_1)$, $\hat\tau\in H^{\frac12}(S)$.
Then there exist functions $\tilde u$ and $\tilde\tau$ such that
\begin{gather}
\tilde u\in H^1(\Omega)^n, \qquad \tilde u\Big\vert_{S_1}=\hat u, \qquad 
        \diver\tilde u=0,     \label{2.7}\\
\tilde\tau\in H^1(\Omega), \qquad \tilde\tau\Big\vert_S=\hat\tau.
                              \label{2.8}
\end{gather}
Suppose also 
\begin{equation}\label{2.9}
K=(K_1,\dots,K_n)\in L_2(\Omega)^n, \qquad F\in L_2(S_2)^n.
\end{equation}

\section{$P$-generalized solution of the coupled problem for the function 
$\phi_2$, and the existence theorem.}

We consider coupled problem for the function $\phi_2$ defined by \eqref{1.12}.
We assume that the functions $e$ and $\psi_1$ satisfy the following conditions:
\begin{description}
\item[(C1)]
$e:(y_1,y_2,x)\to e(y_1,y_2,x)$ is a function continuous in $\R_+\times\R
\times\overline\Omega$ and in addition
\begin{equation}\label{3.1}
0\le e(y_1,y_2,x)\le a_0, \quad (y_1,y_2,x)\in \R_+\times\R\times\overline
\Omega,
\end{equation}
$a_0$ being a positive constant.
\end{description}
\begin{description}
\item[(C2)]
$\psi_1:y_1,y_2,y_3,x\to\psi_1(y_1,y_2,y_3,x)$ is a function continuous in 
$\R_+\times\R_+\times\R\times\overline\Omega$, and for an arbitrary fixed
$(y_2,y_3,x)\in\R_+\times\R\times\overline\Omega$ the function $\psi_1(.,y_2,
y_3,x):y_1\to\psi_1(y_1,y_2,y_3,x)$ is continuously differentiable in $\R_+$,
and the following inequalities hold:
\begin{gather}
a_2\ge\psi_1(y_1,y_2,y_3,x)\ge a_1,  \label{3.2}\\
\psi_1(y_1,y_2,y_3,x)+2\,\frac{\di\psi_1}{\di y_1}\,(y_1,y_2,y_3,x)y_1\ge a_3,
                                     \label{3.3}\\
\Big\vert \frac{\di\psi_1}{\di y_1}(y_1,y_2,y_3,x)\Big\vert y_1\le a_4,
                                     \label{3.4}
\end{gather}
where $a_i$, $1\le i\le 4$, are positive constants.
\end{description}

The inequalities \eqref{3.1} and \eqref{3.2} indicate that the viscosity 
function is 
bounded from below and above by positive constants. \eqref{3.3} implies that 
for fixed values of $\vert E\vert$, $\tau$, $x$ the derivative of the 
function $I(u)\to G(u)$ is positive, where $G(u)$ is the second invariant of
the stress deviator $G(u)=4\phi_2^2\,I(u)$. This means that in the case of 
simple 
shear flow the shear stress increases with increasing shear rate. \eqref{3.4} 
is a restriction on $\frac{\di\psi_1}{\di y_1}$ for large values of $y_1$. The
inequalities \eqref{3.2}--\eqref{3.4} are natural from the physical point of 
view.

We consider the following spaces:
\begin{align}
&X=\{u\vert u\in H^1(\Omega)^n,\quad    u\vert_{S_1}=0\}, \label{3.5}\\
&V=\{u\vert u\in X, \quad\diver u=0\}.                    \label{3.6}
\end{align}
By means of Korn's inequality, the expression
\begin{equation}\label{3.7}
\|u\|_X=\Big(\int_\Omega I(u)\,dx\Big)^{\frac12}
\end{equation}
defines a norm on $X$ and $V$ being equivalent to the norm of $H^1(\Omega)^n$.
 
Everywhere below we use the following notations: if $Y$ is a normed space, we
denote by $Y^*$ the dual of $Y$, and by $(f,h)$ the duality between $Y^*$ and
$Y$, where $f\in Y^*$, $h\in Y$. In particular, if $f\in L_2(\Omega)$ or
$f\in L_2(\Omega)^n$, then $(f,h)$ is the scalar product in $L_2(\Omega)$ or 
in $L_2(\Omega)^n$, respectively. The sign$\rightharpoonup$ denotes weak 
convergence in a Banach space.

Define a bilinear form $\pi$ in $H^1(\Omega)\times H^1(\Omega)$ as follows:
\begin{equation}\label{3.8}
\pi(\theta_1,\theta_2)=\int_\Omega\,\sum_{i=1}^n\,\frac{\di\theta_1}{\di x_i}
\,\frac{\di\theta_2}{\di x_i}\,dx, \quad \theta_1,\theta_2\in H^1(\Omega).
\end{equation}
The expression
\begin{equation}\label{3.9}
\|\theta\|_1=(\pi(\theta,\theta))^{\frac12}
\end{equation}
defines a norm in $H_0^1(\Omega)$ that is equivalent to the norm of $H^1
(\Omega)$.

We introduce a mapping $N:X\times H_0^1(\Omega)\to X^*$ by
\begin{gather}
(N(v,\zeta),h)=2\int_\Omega[e(\vert E\vert,\tilde\tau+\zeta,x)(\lambda+I
(\tilde u+v))^{-\frac12}                            \notag\\
+\psi_1(I(\tilde u+v),\vert E\vert,\tilde\tau+
\zeta,x)]\eps_{ij}(\tilde u+v)\eps_{ij}(h)dx, \qquad
v,h\in X, \quad \zeta\in H_0^1(\Omega), \label{3.10}
\end{gather}
and we assume that $\lambda>0$ in \eqref{3.10}.

Determine an operator $A:X\times H_0^1(\Omega)\to H^{-1}(\Omega)$, where
$H^{-1}(\Omega)$ is the dual space of $H_0^1(\Omega)$, as follows:
\begin{gather}
\Big(A(v,\zeta),\xi\Big)=\int_\Omega(\tilde\tau+\zeta)(\tilde u_i+v_i)
\frac{\di\xi}{\di x_i}\,dx        
   +2\eps\int_\Omega[e(\vert E\vert,\tilde\tau+\zeta,x)(\lambda+
    I(\tilde u+v))^{-\frac12}    \notag\\
   +\psi_1(I(\tilde u+v),\vert E\vert,\tilde\tau+\zeta,x)]\,I(P(\tilde u
   +v))\xi\,dx-\chi\pi(\tilde\tau,\xi), \quad v\in X, \quad (\zeta,\xi)\in
    H_0^1(\Omega)^2. \label{3.11}
\end{gather} 
Here $P$ is an operator of regularization given by
\begin{equation}\label{3.12}
Pu(x)=\int_{\R^n} \omega(\vert x-x'\vert)u(x')dx',\qquad x\in\overline\Omega,
\end{equation}
where
\begin{align}
&\omega\in C^\infty(\R_+),\quad \mbox{supp }\omega=[0,a],\quad \omega(z)\ge 0 
\quad z\in\R_+, \notag\\
&\int_{\R^n}\omega(\vert x\vert)dx=1,\quad a \mbox{ is a small positive
constant}.\label{3.13} 
\end{align}
In \eqref{3.12} we assume that the function $u$ is extended to $\R^n$.

We denote by $B$ the operator $\diver$, i.e.
\begin{equation}\label{3.14}
Bu=\diver u.
\end{equation}
It is obvious that $B\in\cal L (X,L_2(\Omega))$, and we denote by $B^*$ the
adjoint to $B$ operator.

Consider the following problem: find a triple $(v,p,\zeta)$ such that
\begin{gather}
(v,p,\zeta)\in X\times L_2(\Omega)\times H_0^1(\Omega), \label{3.15}\\ 
\Big(N(v,\zeta),h\Big)-\Big(B^*p,h\Big)=\Big(K+F,h\Big), \qquad h\in X,
                                                         \label{3.16}\\
\Big(B\,v,q\Big)=0, \qquad q\in L_2(\Omega),              \label{3.17}\\  
\pi(\zeta,\xi)-\frac1\chi\Big(A(v,\zeta),\xi\Big)=0, \qquad \xi\in H_0^1
(\Omega).                                                \label{3.18}  
\end{gather}
Here we use the notations
\begin{equation}\label{3.19}
\Big(K,h\Big)=\int_\Omega K_ih_i\,dx,\quad \Big(F,h\Big)=\int_{S_2}F_i h_i
ds, \quad h\in X.
\end{equation}
The triple $(u=\tilde u+v,\,p,\,\tau=\tilde\tau+\zeta)$, where 
$v,p,\zeta$ is a solution of the problem \eqref{3.15}--\eqref{3.18},  
will be called the $P$-generalized solution of 
the problem \eqref{2.1}--\eqref{2.6} for the viscosity function $\phi_2$; in 
the special case that $P$ is a unit 
operator the triple $(u=\tilde u+v,\,p,\,\tau=\tilde\tau+\zeta)$
is a generalized solution of the problem \eqref{2.1}--\eqref{2.6}.

Indeed, by use of Green's formula it can be seen that smooth generalized
solution of the problem \eqref{2.1}--\eqref{2.6} is a classical solution of 
this problem. 
On the contrary, if $(u,p,\theta)$ is a classical solution of the problem 
\eqref{2.1}--\eqref{2.6}, then the triple $(v=u-\tilde u,\,p,\,\zeta=\tau
-\tilde\tau)$ is a solution of the problem \eqref{3.15}--\eqref{3.18}, 
wherein $P$ is the unit operator.

From the physical point of view the presence of the operator $P$ in \eqref{3.18}
means that the value of dissipation of energy at a point $x$ depends on the 
values of the rate of strain tensor at points belonging to some small vicinity
of the point $x$. Therefore, the presence of the operator $P$ in \eqref{3.18}
is natural from the physical point of view and it leads to a nonlocal model.

\begin{th}
Let $\Omega$ be a bounded domain in $\R^n$, $n=2$ or $3$ with a Lipschitz 
continuous boundary $S$, and suppose that the conditions $(C1)$, $(C2)$, 
\eqref{2.7}--\eqref{2.9} are satisfied. Then there exists a solution of the 
problem \eqref{3.15}--\eqref{3.18}.
\end{th}

\section{Auxiliary results.}

The next lemma follows from the results of the work \cite{5}.
\begin{lem}
Assume that the conditions $(C1)$, $(C2)$, \eqref{2.7}, \eqref{2.8} are 
satisfied. Then for an arbitrary fixed $\zeta\in H_0^1(\Omega)$ the partial
function $N(.,\zeta):v\to N(v,\zeta)$, where $N(v,\zeta)$ is given by 
\eqref{3.10}, is a Lipschitz-continuous strictly monotone and coercive mapping
of $X$ into $X^*$, i.e.
\begin{gather}
\|N(v,\zeta)-N(w,\zeta)\|_{X^*}\le\mu_1\|v-w\|_X, \qquad v,w\in X,
                                                              \label{4.1}\\
(N(v,\zeta)-N(w,\zeta),v-w)\ge\mu_2\|v-w\|_X^2, \qquad v,w\in X,
                                                              \label{4.2}\\ 
(N(v,\zeta),v)\ge\mu_3\|v\|_X^2-\mu_4\|v\|_X, \qquad v\in X,  \label{4.3}   
\end{gather}
where 
\begin{align}
&\mu_1=2a_2+4(a_4+a_0\lambda^{-\frac12}), \qquad \mu_2=\min(2a_1,2a_3), \notag\\
&\mu_3=2a_1, \qquad \mu_4=2(a_0\lambda^{-\frac12}+a_2)\Big(\int_\Omega I
(\tilde u)\,dx\Big)^{\frac12}.     \label{4.4}
\end{align}
\end{lem}
    \begin{lem}
    Let $\Omega$ be a bounded domain in $\R^n$, $n=2$ or $3$ with a Lipschitz
    continuous boundary $S$, and let the operator $B\in\cal L(X,L_2(\Omega))$ 
    be defined by \eqref{3.14}
    Then, the $\inf$-$\sup$ condition
    \begin{equation}\label{4.5}
    \inf_{\mu\in L_2(\Omega)}\,\,\sup_{v\in X}\,\,\frac{(Bv,\mu)}{\|v\|_X\,
    \|\mu\|_{L_2(\Omega)}}\,\,\ge \beta_1>0
    \end{equation}
    holds true. The operator $B$ is an isomorphism from $V^\bot$  onto $L_2
    (\Omega)$, where $V^\bot$ is orthogonal complement of $V$ in $X$, and the 
    operator $B^*$ that is     adjoint to $B$, is an isomorphism from $L_2
    (\Omega)$ onto the polar  set 
    \begin{equation}\label{4.6}
    V^0=\{f\in X^*,\,(f,u)=0, \quad u\in V\}.
    \end{equation}
    Moreover,
    \begin{gather}
    \|B^{-1}\|_{\cal L(L_2(\Omega),V^\bot)}\le \frac1{\beta_1}, \label{4.7}\\
    \|(B^*)^{-1}\|_{\cal L(V^0,L_2(\Omega))}\le \frac1{\beta_1}.\label{4.8}
    \end{gather}
    \end{lem}
    For a proof see \cite{1}. Lemma 4.2 is a generalization of the 
    inf-sup condition in case that the operator div acts in the subspace 
    $H^1_0(\Omega)^n$ (see \cite{3}). This result was first established in an 
    equivalent form by Ladyzhenskaya and Solonnikov in \cite{6}.

\begin{th}
Suppose the conditions $(C1)$, $(C2)$, \eqref{2.7}--\eqref{2.9} are satisfied.
Then for an arbitrary $\zeta\in H_0^1(\Omega)$ there exists a unique 
pair $(v(\zeta),\,p(\zeta))$ such that 
\begin{gather}
v(\zeta)\in X, \qquad p(\zeta)\in L_2(\Omega),    \label{4.9}\\
\big(N(v(\zeta),\zeta),h\big)-\big(B^*\,p(\zeta),h\big)=\big(K+F,h\big),
  \quad h\in X,                                     \label{4.10}\\
\big(Bv(\zeta),q\big)=0, \qquad q\in L_2(\Omega).  \label{4.11}
\end{gather}
In addition
\begin{equation}\label{4.12}
\|v(\zeta)\|_X\le\frac1{\mu_3}\,\big(\|K+F\|_{V^*}+\mu_4\big), \qquad
     \zeta\in H_0^1(\Omega),
\end{equation}
and the function $\zeta\to(v(\zeta),\,p(\zeta))$ is a compact mapping of
$H_0^1(\Omega)$ into $X\times L_2(\Omega)$. 

The condition
\begin{equation}\notag
\zeta_k\rightharpoonup\zeta_0 \quad \text{in } H_0^1(\Omega) 
\end{equation}
implies 
\begin{align}
&v(\zeta_k)\to v(\zeta_0) \quad \text{in } X, \notag\\
&p(\zeta_k)\to p(\zeta_0) \quad \text{in } L_2(\Omega). \notag
\end{align}
\end{th}
{\bf Proof}. It follows from \eqref{4.10}, \eqref{4.11} and \eqref{3.6} that 
the function $v(\zeta)$ is a solution of the following problem:
\begin{equation}\label{4.13}
v(\zeta)\in V, \qquad \big(N(v(\zeta),\zeta),h\big)=\big(K+F,h\big), \qquad
h\in V.
\end{equation}
Lemma 4.1 and the results on solvability of equations with monotone operators,
see e.g. \cite{7} imply that there exists a unique solution of the problem
\eqref{4.13}. Equality \eqref{4.13} yields
\begin{equation}\label{4.14}
N(v(\zeta),\zeta)-K-F\in V^0.
\end{equation}
By virtue of Lemma 4.2 and \eqref{4.14} there exists a unique function 
$p(\zeta)\in L_2(\Omega)$ such that the pair $(v(\zeta),\,p(\zeta))$ is a
solution of the problem \eqref{4.9}--\eqref{4.11}. 

Taking $h=v(\zeta)$ in \eqref{4.10}, we obtain
\begin{equation}\label{4.15}
(N(v(\zeta),\zeta),v(\zeta))=(K+F,v(\zeta))\le\|K+F\|_{V^*}\|v(\zeta)\|_X,
\end{equation}
and \eqref{4.12} follows from \eqref{4.3} and \eqref{4.15}. Inequality \eqref
{4.12} implies
\begin{equation}\label{4.16}
\|p(\zeta)\|_{L_2(\Omega)}\le c, \qquad \zeta\in H_0^1(\Omega).
\end{equation}
Let now
\begin{equation}\label{4.17}
\zeta_k\rightharpoonup \zeta_0 \text{ in } H_0^1(\Omega).
\end{equation}
We take the following notations:
\begin{gather}
v^m=v(\zeta_m), \quad I_m=I(\tilde u+v^m), \quad \eps_{ij}^m=\eps_{ij}(\tilde
    u+v^m), \quad m=0,1,2,\dots,                    \notag\\
\phi_{mk}=e(\vert E\vert,\tilde\tau+\zeta_k,x)(\lambda+I(\tilde u+v^m))^
          {-\frac12}                                \notag\\
+\psi_1(I(\tilde u+v^m),\vert E\vert,\tilde\tau+\zeta_k,x), \qquad     
          k,m=0,1,2,\dots.                          \label{4.18}
\end{gather}
It follows from \eqref{4.13} that
\begin{align}
&(N(v^k,\zeta_k),h)=(K+F,h), \qquad h\in V, \quad k=1,2,\dots \notag\\
&(N(v^0),\zeta_0),h)=(K+F,h), \qquad h\in V. \notag
\end{align}
Taking  $h=v^k-v^0$, we obtain from here by \eqref{3.10} that
\begin{equation}\label{4.19}
\int_\Omega(\phi_{kk}\,\eps_{ij}^k-\phi_{00}\,\eps_{ij}^0)
      (\eps_{ij}^k-\eps_{ij}^0)\,dx=0.
\end{equation}
\eqref{4.19} implies
\begin{equation}\label{4.20}
M_1^k+M_2^k=0,
\end{equation}
where
\begin{align}
&M_1^k=\int_\Omega(\phi_{kk}\,\eps_{ij}^k-\phi_{0k}\,\eps_{ij}^0)
       (\eps_{ij}^k-\eps_{ij}^0)\,dx,                    \label{4.21}\\
&M_2^k=\int_\Omega(\phi_{0k}-\phi_{00})\,\eps_{ij}^0
       (\eps_{ij}^k-\eps_{ij}^0)\,dx.                    \label{4.22}
\end{align}
The Cauchy inequality and the notation \eqref{4.18} imply
\begin{equation}\label{4.23}
\vert M_2^k\vert\le \Big(\int_\Omega(\phi_{0k}-\phi_{00})^2 I_0\,dx\Big)^
{\frac12}\Big(\int_\Omega I(v^k-v^0)\,dx\Big)^{\frac12}.
\end{equation}
By virtue of \eqref{4.17}, we can extract a subsequence $\{\zeta_\eta\}$ from
the sequence $\{\zeta_k\}$ such that
\begin{equation}\label{4.24}
\zeta_\eta\to\zeta_0 \quad \text{a.e. in } \Omega.
\end{equation}
Now the continuity of the functions $e$ and $\psi_1$ (see (C1), (C2)) yields
\begin{equation}\notag
\phi_{0\eta}\to\phi_{00} \quad \text{a.e. in } \Omega.
\end{equation}
The functions $\phi_{0\eta}$ are bounded by a constant $a_0\lambda^{-\frac12}+
a_2$. Therefore, the Lebesque theorem and \eqref{4.23} imply that
\begin{equation}\label{4.25}
\lim M_2^\eta=0.
\end{equation}
By analogy with the stated above, we can see that from any subsequence $\{M_2
^m\}$ extracted from the sequence $\{M_2^k\}$, one can extract a subsequence
$\{M_2^{m_i}\}$ such that $\lim M_2^{m_i}=0$. Therefore, $\lim M_2^k=0$, and by
\eqref{4.20}, we get
\begin{equation}\label{4.26}
\lim M_1^k=0.
\end{equation}
It follows from \eqref{3.10}, \eqref{4.2}, \eqref{4.18} and \eqref{4.21} that
\begin{equation}\notag
M_1^k=\frac12\,\Big[(N(v^k,\,\zeta_k)-N(v^0,\zeta_k),v^k-v^0)\Big]\ge\frac12
                  \,\mu_2\|v^k-v^0\|_X^2.
\end{equation}
This inequality together with \eqref{4.26} yields
\begin{equation}\label{4.27}
v(\zeta_k)\to v(\zeta_0) \quad \text{in } X.
\end{equation}
We have
\begin{gather}
\|N(v(\zeta_k),\zeta_k)-N(v(\zeta_0),\zeta_0)\|_{X^*}    \notag\\
\le\|N(v(\zeta_k),\zeta_k)-N(v(\zeta_0),\zeta_k)\|_{X^*}+\|N(v(\zeta_0),
\zeta_k)-N(v(\zeta_0),\zeta_0)\|_{X^*}.  \notag
\end{gather}
By virtue of \eqref{4.1} and \eqref{4.27} the first term on the right-hand
side of this inequality tends to zero, and \eqref{4.24} implies that the 
second term also tends to zero.

Therefore,
\begin{equation}\label{4.28}
N(v(\zeta_k),\zeta_k)\to N(v(\zeta_0),\zeta_0) \quad \text{in } X^*.
\end{equation}
It follows from \eqref{4.10} that
\begin{equation}\label{4.29}
B^*\,p(\zeta_k)-B^*\,p(\zeta_0)=N(v(\zeta_k),\zeta_k)-N(v(\zeta_0),\zeta_0)
\quad \text{in }X^*.
\end{equation}
Since the operator $B^*$ is an isomorphism from $L_2(\Omega)$ onto $V^0$, we
obtain from \eqref{4.28} and \eqref{4.29} that
\begin{equation}\notag
p(\zeta_k)\to p(\zeta_0) \quad \text{in } L_2(\Omega),
\end{equation}
and the theorem is proved.
$\blacksquare$

We assign a mapping $G:H_0^1(\Omega)\to H^{-1}(\Omega)$ as follows:
\begin{equation}\notag
H_0^1(\Omega)\ni\zeta\to G(\zeta)=A(v(\zeta),\zeta)\in H^{-1}(\Omega).
\end{equation}
Here $A:X\times H_0^1(\Omega)\to H^{-1}(\Omega)$ is given by \eqref{3.11} and
$v(\zeta)$ is the solution of the problem \eqref{4.9}--\eqref{4.11}.

\begin{lem}
Suppose the conditions $(C1)$, $(C2)$, \eqref{2.7}--\eqref{2.9} are satisfied.
Then $G$ is a compact mapping of $H_0^1(\Omega)$ into $H^{-1}(\Omega)$, 
the condition $\zeta_k\rightharpoonup\zeta_0$ in $H_0^1(\Omega)$ implies
$G(\zeta_k)\to G(\zeta_0)$ in $H^{-1}(\Omega)$.
\end{lem}
{\bf Proof}. Let $\zeta_k\rightharpoonup\zeta_0$ in $H_0^1(\Omega)$. Then 
Theorem 4.1 and the embedding theorem yield
\begin{align}
&v(\zeta_k)\to v(\zeta_0) \quad \text{in } L_4(\Omega)^n,      
                                                    \label{4.30}\\
&\eps_{ij}(v(\zeta_k))\to\eps_{ij}(v(\zeta_0)) \quad \text{in } L_2(\Omega),
                        \quad i,j=1,\dots,n,        \label{4.31}\\
&\zeta_k\to\zeta_0 \quad \text{in } L_4(\Omega).  \label{4.32}
\end{align}
In view of \eqref{4.31}, \eqref{4.32} a subsequence $\{\zeta_\eta\}$ can be
extracted from the sequence $\{\zeta_k\}$ such that
\begin{align}
&\zeta_\eta\to\zeta_0 \quad \text{a.e. in } \Omega,      \label{4.33}\\
&I(v(\zeta_\eta))\to I(v(\zeta_0)) \quad \text{a.e. in } \Omega. \label{4.34}
\end{align}
We represent the mapping $G$ as a sum of two mappings $G=G_1+G_2$, which we
define as follows:
\begin{align}
(G_1(\zeta),\xi)&=\int_\Omega(\tilde\tau+\zeta)(\tilde u_i+v(\zeta)_i)
                   \,\frac{\di\xi}{\di x_i}\,dx-\chi\pi(\tilde\tau,\xi), 
                   \quad \zeta,\xi\in H_0^1(\Omega),        \label{4.35}\\
(G_2(\zeta),\xi)&=2\eps\int_\Omega[e(\vert E\vert,\tilde\tau+\zeta,x)
                   (\lambda+I(\tilde u+v(\zeta))^{-\frac12}  \notag\\
                 &+\psi_1(I(\tilde u+v(\zeta)),\vert E\vert,\tilde\tau+
                   \zeta,x)]\,I(P(\tilde u+v(\zeta))\,\xi\,dx, \quad 
                   \zeta,\xi\in H_0^1(\Omega)                \label{4.36}
\end{align}
If we will argue that
\begin{align}
&G_1(\zeta_k)\to G_1(\zeta_0) \quad \text{in } H^{-1}(\Omega), \label{4.37}\\
&G_2(\zeta_k)\to G_2(\zeta_0) \quad \text{in } H^{-1}(\Omega), \label{4.38}
\end{align}
then the lemma will be proved.

We denote
\begin{equation}\label{4.39}
f_{ki}=(\tilde\tau+\zeta_k)(\tilde u_i+v(\zeta_k)_i)-(\tilde\tau+
\zeta_0)(\tilde u_i+v(\zeta_0)_i).            
\end{equation}
It follows from \eqref{4.35} that
\begin{equation}\label{4.40}
\Big\vert(G_1(\zeta_k)-G_1(\zeta_0),\,\xi)\Big\vert=\Big\vert\int_\Omega 
f_{ki}\,\frac{\di\xi}{\di x_i}\,dx\Big\vert                   
\le\Big(\sum_{i=1}^n\int_\Omega f_{ki}^2\,dx\Big)^{\frac12}\|\xi\|_1,
\end{equation}
$\|\cdot\|_1$ being the norm defined by \eqref{3.9}.

\eqref{4.30} and \eqref{4.32} yield
\begin{equation}\label{4.41}
f_{ki}\to 0 \qquad \text{in } L_2(\Omega)\quad \text{as } k\to\infty, \quad
                   i=1,\dots,n.
\end{equation}
By \eqref{4.40} and \eqref{4.41}, we obtain \eqref{4.37}.

By applying \eqref{3.1}, \eqref{3.2}, the H\"older inequality, and the 
notations \eqref{4.18}, we obtain
\begin{gather}
\Big\vert\frac1{2\eps}(G_2(\zeta_k)-G_2(\zeta_0),\xi)\Big\vert=\Big\vert
\int_\Omega[\phi_{kk}\,I(P(\tilde u+v^k))                 \notag\\ 
-\phi_{00}\,I(P(\tilde u+v^0))]\xi\,dx\Big\vert=\Big\vert\int_\Omega[\phi_{kk}
(P\eps_{ij}^k-P\eps_{ij}^0)(P\eps_{ij}^k+P\eps_{ij}^0)    \notag\\
+(\phi_{kk}-\phi_{00})\,I(P(\tilde u+v^0))]\,\xi\,dx\Big\vert
\le(\delta\alpha_{1k}+\alpha_{2k})\|\xi\|_{L_6(\Omega)}\le c(\delta\alpha_{1k}
+\alpha_{2k})\|\xi\|_1,                                    \notag\\
\xi\in H_0^1(\Omega).    \label{4.42}
\end{gather}
Here
\begin{align}
\delta&=a_0\,\lambda^{-\frac12}+a_2,   \label{4.43}\\
\alpha_{1k}&=\sum_{i,j=1}^n\,(\|P(\eps_{ij}^k-\eps_{ij}^0)\|_{L_{\frac{12}5}
(\Omega)}\|P(\eps_{ij}^k+\eps_{ij}^0)\|_{L_{\frac{12}5}(\Omega)},   
                                             \label{4.44}\\
\alpha_{2k}&=\Big(\int_\Omega(\phi_{kk}-\phi_{00})^{\frac65}\,(I(P(\tilde u
+v^0)))^{\frac65}\,dx\Big)^{\frac56}.        \label{4.45}
\end{align}
\eqref{4.31} yields
\begin{equation}\label{4.46}
\lim\alpha_{1k}=0. 
\end{equation}
By \eqref{4.33}, \eqref{4.34} and the Lebesgue theorem, we obtain that 
$\lim\alpha_{2\eta}=0$.
By analogy with the stated above, we can see that from any subsequence 
$\{\zeta_m\}$ extracted from the sequence $\{\zeta_k\}$, one can  extract a
subsequence $\{\zeta_i\}$ such that $\lim\alpha_{2i}=0$. Therefore,
\begin{equation}\label{4.47}
\lim\alpha_{2k}=0.
\end{equation}
\eqref{4.42}, \eqref{4.46} and \eqref{4.47} imply \eqref{4.38}, and the lemma 
is proved.
$\blacksquare$

\section{Proof of the Theorem 3.1.}

Consider the problem: find a function $\zeta$ such that 
\begin{gather}
\zeta\in H_0^1(\Omega), \label{5.1}\\
\pi(\zeta,\xi)-\frac1{\chi}(A(v(\zeta),\zeta),\xi)=0, \qquad \xi\in H_0^1
(\Omega), \label{5.2}
\end{gather}
where $v(\zeta)$ is the solution of the problem \eqref{4.13}. In this case 
there exists a unique function $p(\zeta)\in L_2(\Omega)$ such that the pair
$(v(\zeta),\,p(\zeta))$ is the solution of the problem \eqref{4.9}--\eqref{4.11}.

It is obvious that, if $\zeta$  is a solution of the problem \eqref{5.1}, 
\eqref{5.2}, then the triple $(v=v(\zeta),\,p=p(\zeta),\zeta)$ is a solution
of the problem \eqref{3.15}--\eqref{3.18}.

Let $\Big\{\cal Z_k\Big\}$ be a sequence of finite dimensional subspaces in
$H_0^1(\Omega)$, such that
\begin{equation}\label{5.2a}
\cal Z_k\subset\cal Z_{k+1},\qquad \lim_{k\to\infty}\,\inf_{h\in\cal Z_k}\,\|
         w-h\|_1=0, \qquad w\in H_0^1(\Omega).
\end{equation}
We seek an approximate solution of the problem \eqref{5.1}, \eqref{5.2} of the
form
\begin{equation}\label{5.3}
\zeta_k\in\cal Z_k, \quad \pi(\zeta_k,\xi)-\frac1{\chi}(A(v(\zeta_k),
\zeta_k),\xi)=0, \quad \xi\in\cal Z_k,
\end{equation}
where $v(\zeta_k)$ is the solution of the problem \eqref{4.13} for $\zeta=
\zeta_k$.

Applying the integration by parts, we obtain
\begin{gather}
\int_\Omega \zeta(\tilde u_i+v_i)\frac{\di\zeta}{\di x_i}\,dx=-\int_\Omega
\frac{\di\zeta}{\di x_i}(\tilde u_i+v_i)\zeta\,dx=0, \notag\\
v\in V, \qquad \zeta\in H_0^1(\Omega).                \label{5.4}
\end{gather}
Taking into account \eqref{2.7}, \eqref{2.8}, \eqref{3.1}, \eqref{3.2},
\eqref{4.12} and \eqref{5.4}, we reduce from \eqref{3.11} that 
\begin{equation}\label{5.5}
\Big\vert\frac1{\chi}\Big(A(v(\zeta),\zeta),\zeta\Big)\Big\vert\le c
\|\zeta\|_1, \qquad \zeta\in H_0^1(\Omega).
\end{equation}
Therefore,
\begin{equation}\label{5.6}
y(\zeta)=\pi(\zeta,\zeta)-\frac1{\chi}(A(v(\zeta),\zeta),\zeta)\ge
\|\zeta\|_1^2-c\|\zeta\|_1,  
\end{equation}
and $y(\zeta)\ge 0$ for $\|\zeta\|_1\ge c$.

From the corollary of Brouwer's fixed point theorem \cite{2} it follows that 
there exists a solution of the problem \eqref{5.3} and $\|\zeta_k\|_1\le c$.
Consequently, we can extract a subsequence $\{\zeta_\eta\}$ from the sequence
$\{\zeta_k\}$ such that
\begin{equation}\label{5.7}
\zeta_\eta\rightharpoonup\zeta_0 \text{ in } H_0^1(\Omega).
\end{equation}
Let $\eta_0$ be a fixed positive number and $\xi\in\cal Z_{\eta_0}$. By 
\eqref{5.7} and Lemma 4.3, we pass to the limit in \eqref{5.3} with $k$ 
replaced by $\eta$. Then, we get
\begin{equation}\label{5.8}
\zeta_0\in H_0^1(\Omega), \quad \pi(\zeta_0,\xi)-\frac1{\chi}(A(v(\zeta_0),
\zeta_0),\xi)=0, \quad \xi\in\cal Z_{\eta_0}.
\end{equation}
\eqref{5.2a} and \eqref{5.8} imply that the function $\zeta=\zeta_0$ is a 
solution of the problem \eqref{5.1}, \eqref{5.2}. The theorem is proved.
$\blacksquare$

\section{Generalized solution of the coupled problem for the function 
$\phi_2$.}

We denote
\begin{equation}\notag
Q=H_0^1(\Omega)\cap L_\infty(\Omega),
\end{equation}
and determine a mapping $A_1:X\times W_0^{1,\frac65}(\Omega)\to Q^*$ as
follows:
\begin{gather}
(A_1(v,\zeta),\xi)=-\int_\Omega\,\frac{\di(\tilde\tau+\zeta)}{\di x_i}(\tilde
              u_i+v_i)\xi\,dx                        \notag\\
             +2\eps\int_\Omega[e(\vert E\vert,\tilde\tau+\zeta,x)(\lambda
             +I(\tilde u+v))^{-\frac12}             \notag\\
             +\psi_1(I(\tilde u+v),\vert E\vert,\tilde\tau+\zeta,x)]\,I(\tilde 
              u+v))\xi\,dx-\chi\pi(\tilde\tau,\xi), \notag\\
              v\in X, \qquad \zeta\in W_0^{1,\frac65}(\Omega),\qquad\xi\in Q.
                                                     \label{6.1} 
\end{gather}
Here $W_0^{1,\frac65}(\Omega)$ is the closure of $D(\Omega)$ for the norm of
the Sobolev space $W^{1,\frac65}(\Omega)$.

The expression
\begin{equation}\label{6.2}
\|\zeta\|_2=\Big(\int_\Omega\sum_{i=1}^n\Big\vert\frac{\di\zeta}{\di x_i}
      \Big\vert^{\frac65}\,\,dx\Big)^{\frac56},
\end{equation}
defines a norm in $W_0^{1,\frac65}(\Omega)$ which is equivalent to the norm of
$W^{1,\frac65}(\Omega)$.

Consider the problem: find a triple $(v,p,\zeta)$ such that
\begin{align}
&(v,p,\zeta)\in X\times L_2(\Omega)\times W_0^{1,\frac65}(\Omega), 
                                                       \label{6.3}\\
&(N(v,\zeta),h)-(B^*\,p,h)=(K+F,h), \qquad h\in X,     \label{6.4}\\ 
&(Bv,q)=0, \qquad q\in L_2(\Omega),                    \label{6.5}\\
&\pi(\zeta,\xi)-\frac1\chi\,(A_1,(v,\zeta),\xi)=0, \qquad \xi\in Q.
                                                       \label{6.6}
\end{align}
The triple $(u=\tilde u+v,p,\tau=\tilde\tau+\zeta)$, where $(v,p,\xi)$ is a
solution of the problem \eqref{6.3}--\eqref{6.6}, is a generalized solution
of the problem \eqref{2.1}--\eqref{2.6}.

Indeed, by use of the Green formula, one can verify that, if $(v,p,\zeta)$ 
is a solution of the problem \eqref{6.3}--\eqref{6.6}, then $u,p,\tau$ with
$u=\tilde u+v$, and $\tau=\tilde\tau+\zeta$ is a solution of the problem
\eqref{2.1}--\eqref{2.6} in the distribution sense. On the contrary, if $u,p,
\tau$ is a solution of the problem \eqref{2.1}--\eqref{2.6} such that 
\eqref{6.3} holds with $v=u-\tilde u$, $\zeta=\tau-\tilde\tau$, then
$(v,p,\zeta)$ is a solution of the problem \eqref{6.3}--\eqref{6.6}.

\begin{th}
Let $\Omega$ be a bounded domain in $\R^n$, $n=2$ or $3$ with a boundary of 
the class $C^4$. Suppose the conditions $(C1)$, $(C2)$, \eqref{2.7}--
\eqref{2.9} are satisfied.

Assume also that
\begin{equation}\label{6.7}
\inf_{w\in V}\,\frac{\tilde c}\chi\Big[\|\tilde u+w\|_{H^1(\Omega)^n}\,+
\frac1{2a_1}\Big(\|K+F\|_{V^*}   
       +2(a_0\lambda^{-\frac12}\,+a_2)\Big(\int_\Omega I(\tilde u+w)dx\Big)^
{\frac12}\Big)\Big]< 1,        
\end{equation}
where $\tilde c$ is a constant depending on the domain $\Omega$. Then, there 
exists a solution of the problem \eqref{6.3}--\eqref{6.6}
\end{th}
{\bf Proof.} 

1)Let $\{P_k\}_{k=1}^\infty$ be a sequence of regularizing 
operators assigned by
\begin{equation}\label{6.8}
P_k u(x)=\int_{\R^n}\omega_k(\vert x-x'\vert)u(x')dx', \quad 
                           x\in\overline\Omega,
\end{equation}
where
\begin{equation}\label{6.9}
\omega_k\in C^\infty(\R_+), \text{supp}\,\omega_k=[0,b_k], \quad 
\omega_k(z)\ge 0,\quad z\in\R_+, \quad \int_{\R^n}\omega_k(\vert x\vert)dx=1,    
\end{equation}
and also $\lim\,b_k=0$. In this case we have
\begin{align}
&\lim\|P_k\,u-u\|_{H^1(\Omega)^n}=0, \qquad u\in H^1(\Omega)^n, \notag\\
&\lim\|P_k\,u-u\|_{L_2(\Omega)^n}=0, \qquad u\in L_2(\Omega)^n, \label{6.10}
\end{align}
and it is assumed here that the function $u$ is prolonged in $\Omega_1$,
$\Omega_1\supset\overline\Omega$, such that $u$ belongs to $H^1(\Omega_1)^n$ 
or $L_2(\Omega_1)^n$, respectively.

By virtue of the Theorem 3.1 for each operator $P=P_k$ there exists a solution
of the problem \eqref{3.15}--\eqref{3.18}, which we denote by $v^k$, $p_k$, 
$\zeta_k$. In this case the function $\zeta_k$ is a solution of the following 
problem:
\begin{equation}\label{6.11}
\zeta_k\in H_0^1(\Omega), \quad \sum_{i=1}^n\,\frac{\di^2\zeta_k}{\di x_i^2}=
\beta_k-\sum_{i=1}^n\frac{\di^2\tilde\tau}{\di x_i^2} \quad \text{in } 
         D^*{(\Omega)},
\end{equation}
where
\begin{equation}\label{6.12}
\beta_k=\frac1{\chi}(\tilde u_i+v_i^k)\frac{\di(\tilde\tau+\zeta_k)}{\di x_i}
       -\frac{2\eps}{\chi}\,\phi_{kk}\,I(P_k(\tilde u+v^k)),
\end{equation}
and $\phi_{kk}$ is defined in \eqref{4.18}.

We have
\begin{equation}\label{6.13}
\Big\|(\tilde u_i+v_i^k)\frac{\di(\tilde\tau+\zeta_k)}{\di x_i}\Big\|_
    {L_1(\Omega)}\le M_{1k}+M_{2k},
\end{equation}
where 
\begin{equation}\label{6.14}
M_{1k}=\Big\|(\tilde u_i+v_i^k)\,\frac{\di\tilde\tau}{\di x_i}\Big\|_{L_1
  (\Omega)},\quad M_{2k}=\Big\|(\tilde u_i+v_i^k)\,\frac{\di\zeta_k}{\di x_i}
  \Big\|_{L_1(\Omega)}.
\end{equation}
It follows from \eqref{2.7}--\eqref{2.9}, \eqref{4.4} and \eqref{4.12} that
\begin{equation}\label{6.15}
M_{1k}=\sum_{i=1}^n\|\tilde u_i+v_i^k\|_{L_6(\Omega)}\Big\|\frac{\di\tilde
\tau}{\di x_i}\Big\|_{L_{\frac65}(\Omega)}\le c_1\|\tilde u+v^k\|_{H^1
(\Omega)^n}\|\tilde\tau\|_2\le c_2.
\end{equation}
\eqref{2.7}, \eqref{4.4} and \eqref{4.12} imply
\begin{equation}\label{6.16}
M_{2k}\le\sum_{i=1}^n\|\tilde u_i+v_i^k\|_{L_6(\Omega)}\Big\|\frac{\di\zeta_k}
{\di x_i}\Big\|_{L_{\frac65}(\Omega)}             
   \le c_3(\|\tilde u\|_{H^1(\Omega)^n}+\|v^k\|_X)\|\zeta_k\|_2\le c_3\check c
\|\zeta_k\|_2,  
\end{equation}
where
\begin{equation}\label{6.17}
\check c=\|\tilde u\|_{H^1(\Omega)^n}+\frac1{2a_1}\Big(\|K+F\|_{V^*}+2(a_0
\lambda^{-\frac12}+a_2)\Big(\int_\Omega I(\tilde u)dx\Big)^{\frac12}\Big).
\end{equation}
Taking note of \eqref{6.13}, \eqref{6.15} and \eqref{6.16}, we obtain
\begin{equation}\label{6.18}
\|(\tilde u_i+v_i^k)\,\frac{\di(\tilde\tau+\zeta_k)}{\di x_i}\|_{L_1(\Omega)}
\le c_3\,\check c\|\zeta_k\|_2+c_2.
\end{equation}
\eqref{2.7}, \eqref{3.1}, \eqref{3.2} and  \eqref{4.12} yield
\begin{equation}\label{6.19}
\|\phi_{kk}\,I(P_k(\tilde u+v^k))\|_{L_1(\Omega)}\le c_4.
\end{equation}
It follows from \eqref{6.12}, \eqref{6.18} and \eqref{6.19} that
\begin{equation}\label{6.20}
\|\beta_k\|_{L_1(\Omega)}\le\,\frac{c_3\check c}{\chi}\|\zeta_k\|_2+c_5.
\end{equation}
Since the boundary $S$ of the class $C^4$ and
\begin{equation}\notag
\frac{\di^2\tilde\tau}{\di x_i^2}\in H^{-1}(\Omega)\subset W^{-1,\frac65}
(\Omega), \quad i=1,\dots,n,
\end{equation}
where  $W^{-1,\frac65}(\Omega)$ is the space dual to
$W_0^{1,6}(\Omega)$, we obtain from 
\eqref{6.11} and known results \cite{13}, \cite{14}, Section 1.1 that
\begin{equation}\label{6.21}
\|\zeta_k\|_2\le c_6(\|\beta_k\|_{W^{-1,\frac65}(\Omega)}+\|\tilde\tau\|_
{W^{1,\frac65}(\Omega)}).
\end{equation}
The embedding of $L_1(\Omega)$ into $W^{-1,\frac65}(\Omega)$ is continuous
because $W_0^{1,6}(\Omega)\subset C(\overline\Omega)$ for $n=2$ and 3. So
that by \eqref{6.20} and \eqref{6.21}, we obtain
\begin{equation}\label{6.22}
\|\zeta_k\|_2\le c_7\|\beta_k\|_{L_1(\Omega)}+c_8\le\frac{c_9\check c}{\chi}
\|\zeta_k\|_2+c_{10}.      
\end{equation}
That is
\begin{equation}\label{6.22a}
\Big(1-\frac{c_9\check c}{\chi}\Big)\|\zeta_k\|_2\le c_{10}
\end{equation}

It is evident that, if a function $\tilde u$ satisfies the conditions 
\eqref{2.7}, then the function $\tilde u+w$, where $w\in V$, satisfies 
\eqref{2.7} also. Therefore, we can consider that, in the operators $N,A$ 
and $A_1$ the function $\tilde u$ is replaced by the function $\tilde u+w$, 
where $w$ is a  function from $V$.
In this case, if $(v,p,\zeta)$  is a solution of the problem 
\eqref{6.3}--\eqref{6.6} for the function $\tilde u$, then $(v-w,\,p,\,\zeta)$
is a solution of \eqref{6.3}--\eqref{6.6} for the function $\tilde u+w$, and
instead of \eqref{6.22a}, we obtain the following inequality
\begin{equation}\label{6.23a}
[1-\gamma(w)]\|\zeta_k\|_2\le c_{10},
\end{equation}
where, see \eqref{6.17},
\begin{equation}\label{6.24a}
\gamma(w)=\frac{c_9}\chi\Big[\|\tilde u+w\|_{H^1(\Omega)^n}+\frac 1{2a_1}\Big
(\|K+F\|_{V^*}+2(a_0\lambda^{-\frac12}+a_2)\Big(\int_\Omega I(\tilde u+w)dx
\Big)^{\frac 12}\Big)\Big].
\end{equation}
If the condition \eqref{6.7} with $\tilde c=c_9$ is satisfied, then there 
exists a function $\check w\in V$ such that $\gamma(\check w)<1$, and by 
virtue of \eqref{6.23a} the sequence $\{\zeta_k\}$ is bounded in $W_0^{1,
\frac 65}(\Omega)$.

We consider that the function $\tilde u$ is replaced by the function
$\tilde u+\check w$ in the operators $N,A$ and $A_1$, and we still denote
by $\tilde u$ the function $\tilde u+\check w$. Then for the new function
$\tilde u$, we have in \eqref{6.22a} $c_9\check c\chi^{-1}<1$. Therefore, 
a subsequence $\{v^m,\,p_m,\,\zeta_m\}$ can be extracted from the sequence 
$\{v^k,\,p_k,\,\zeta_k\}$ such that
\begin{align}
&\zeta_m\rightharpoonup\zeta_0 \quad \text{in } W_0^{1,\frac65}(\Omega),
                                                             \label{6.23}\\
&\zeta_m\to\zeta_0 \quad \text{in } L_{\frac65(\Omega)}  \quad 
                        \text{and a.e. in } \Omega,          \label{6.24}\\
&v^m\rightharpoonup v^0  \quad \text{in } X,                 \label{6.25}\\
&N(v^m,\zeta_m)\rightharpoonup \alpha \quad \text{in } X^*,  \label{6.26}\\
&p_m\rightharpoonup p_0 \quad \text{in } L_2(\Omega),        \label{6.27}
\end{align}

2). Now we are concerned with the passage to the limit.
It follows from \eqref{6.4} and \eqref{6.5} that
\begin{align}
&(N(v^m,\zeta_m),h)-(B^*\,p_m,h)=(K+F,h), \quad  h\in X, \label{6.28}\\
&(Bv^m,q)=0, \quad q\in L_2(\Omega).                           \label{6.29}
\end{align}
Observing \eqref{6.25}--\eqref{6.27}, we pass to the limit in \eqref{6.28}, 
\eqref{6.29}, and obtain
\begin{gather}
\alpha-B^*\,p_0=K+F \quad \text{in } X^*,     \label{6.30}\\
\diver v^0=0.                                 \label{6.31}
\end{gather}
Lemma 4.1 implies
\begin{equation}\label{6.32}
(N(v^m,\,\zeta_m)-N(g,\zeta_m),v^m-g)\ge 0, \quad g\in X.
\end{equation}
Taking into account that $(p_m,B\,v_m)=0$, by \eqref{6.25} and \eqref{6.28}, 
we obtain
\begin{equation}\label{6.33}
(N(v^m,\zeta_m),v^m)=(K+F,v^m)\to(K+F,v^0),
\end{equation}
and
\begin{equation}\label{6.34}
\lim(N(v^m,\zeta_m),g)-(B^*\,p_0,g)=(K+F,g), \quad g\in X.
\end{equation}
By using (C1), (C2), \eqref{6.24} and the Lebesgue theorem, we deduce
\begin{gather}
[e(\vert E\vert,\tilde\tau+\zeta_m,x)(\lambda+I(\tilde u+g))^{-\frac12}+
           \psi_1(I(\tilde u+g),\vert E\vert,\tilde\tau+\zeta_m,x)]
                                                        \notag\\
\times\eps_{ij}(\tilde u+g)\to[e(\vert E\vert,\tilde\tau+\zeta_0,x)
           (\lambda+I(\tilde u+g))^{-\frac12}           \notag\\
+\psi_1(I(\tilde u+g),\vert E\vert,\tilde\tau+\zeta_0,x)]\,\eps_{ij}
           (\tilde u+g) \quad \text{in } L_2(\Omega)\quad
           \text{as }    m\to\infty.                 \label{6.35}
\end{gather}
\eqref{6.25} yields
\begin{equation}\label{6.36}
\eps_{ij}(v^m-g)\rightharpoonup\eps_{ij}(v^0-g) \quad \text{in } L_2(\Omega).
\end{equation}
It follows from \eqref{3.10}, \eqref{6.35} and \eqref{6.36} that
\begin{equation}\label{6.37}
(N(g,\zeta_m),v^m-g)\to(N(g,\zeta_0),v^0-g).
\end{equation}
Observing \eqref{6.33}, \eqref{6.34} and \eqref{6.37}, we pass to the limit 
in \eqref{6.32}, this gives
\begin{equation}\label{6.38}
(K+F-N(g,\zeta_0)+B^*\,p_0,v^0-g)\ge 0, \quad g\in X.
\end{equation}
We choose $g=v^0-\gamma\,h$, $h\in X$, $\gamma>0$, and consider $\gamma\to 0$.
Then, \eqref{4.1} implies
\begin{equation}\label{6.39}
(K+F-N(v^0,\zeta_0)+B^*\,p_0,h)\ge 0.
\end{equation}
This inequality holds for any $h\in X$. Replacing $h$ by $-h$ shows that
here the equality holds true. Consequently, the triple $(v=v^0,\,p=p_0,\,
\zeta=\zeta_0)$ meets the equations \eqref{6.4} and \eqref{6.5}.

It follows from \eqref{4.2} that
\begin{equation}\label{6.40}
(N(v^m,\,\zeta_m)-N(v^0,\zeta_m),v^m-v^0)\ge\mu_2\|v^m-v^0\| _X^2.
\end{equation}
Granting \eqref{6.33}, \eqref{6.34}, \eqref{6.37} one can recognize that the
left-hand side of \eqref{6.40} tends to zero. Therefore, 
\begin{equation}\label{6.41}
v^m\to v^0 \quad \text{in } X,
\end{equation}
and we can regard that
\begin{equation}\label{6.42}
I(v^m)\to I(v^0) \quad \text{a.e. in } \Omega.
\end{equation}
Otherwise, we can extract a subsequence, still denoted by $\{v^m\}$, such
that \eqref{6.42} holds true.

3). The pair $(v^m,\zeta_m)$ meets the following equation (see \eqref{3.11}
and \eqref{3.18})
\begin{equation}\label{6.43}
\pi(\zeta_m,\xi)-\frac1{\chi}\,(U_m,\xi)=0, \quad \xi\in H_0^1(\Omega)\cap
            L_\infty(\Omega)=Q,
\end{equation}
where
\begin{equation}\label{6.44}
(U_m,\xi)=-\int_\Omega\frac{\di(\tilde\tau+\zeta_m)}{\di x_i}(\tilde u_i
          +v_i^m)\xi\,dx   
          +2\eps\int_\Omega\phi_{mm}I(P_m(\tilde u+v^m))\xi\,dx-\chi\pi
            (\tilde\tau,\xi), 
\end{equation}
$\phi_{mm}$ being defined in \eqref{4.18}.

By virtue of \eqref{6.41} $v^m\to v^0$ in $L_6(\Omega)^n$, so that by
\eqref{6.23}, we obtain
\begin{equation}\label{6.45}
\lim_{m\to\infty}\int_\Omega\frac{\di(\tilde\tau+\zeta_m)}{\di x_i}\,(\tilde
   u_i+v_i^m)\xi\,dx=\int_\Omega\frac{\di(\tilde\tau+\zeta_0)}{\di x_i}\,
   (\tilde u_i+v_i^0)\xi\,dx, \quad \xi\in Q.
\end{equation}
We have
\begin{equation}\label{6.46}
\Big\vert\int_\Omega[\phi_{mm}I(P_m(\tilde u+v^m))-\phi_{00}I(\tilde u+v^0)]
   \xi\,dx\Big\vert\le\|\xi\|_{L_\infty(\Omega)}\,(\cal M_1^m+\cal M_2^m),
\end{equation}
where
\begin{align}
&\cal M_1^m=\Big\vert\int_\Omega(\phi_{mm}-\phi_{00})I(\tilde u+v^0)dx\Big
\vert,                              \label{6.47}\\
&\cal M_2^m=\Big\vert\int_\Omega\sum_{i,j=1}^n\phi_{mm}(P_m\eps_{ij}^m-
       \eps_{ij}^0)(P_m\eps_{ij}^m+\eps_{ij}^0)dx\Big\vert,   \label{6.48}
\end{align}
$\eps_{ij}^m$ being defined in \eqref{4.18}.

Taking into consideration (C1), (C2), \eqref{6.24} and \eqref{6.42}, we deduce 
from the Lebesgue theorem that                  
\begin{equation}\label{6.49}
\lim\cal M_1^m=0. 
\end{equation}
(C1) and (C2) imply
\begin{align}
\cal M_2^m &\le(a_0\lambda^{-\frac12}+a_2)\sum_{i,j=1}^n\|P_m\eps_{ij}^m
             - \eps_{ij}^0\|_{L_2(\Omega)}                 \notag\\
           &\times\|P_m\eps_{ij}^m+
            \eps_{ij}^0\|_{L_2(\Omega)}       
            \le c\sum_{i,j=1}^n \|P_m\eps_{ij}^m-\eps_{ij}^0\|_{L_2(\Omega)}
                                                             \notag\\
          &\le c\sum_{i,j=1}^n(\|P_m(\eps_{ij}^m-\eps_{ij}^0)\|_{L_2(\Omega)}
             +\|P_m\eps_{ij}^0-\eps_{ij}^0\|_{L_2(\Omega)})
                                                       \label{6.50}
\end{align}
It follows from \eqref{6.10} and the Banach-Steinhaus theorem, that the norms 
of the operators $P_m$ are uniformly bounded in $H^1(\Omega)^n$ and
$L_2(\Omega)$. So that \eqref{6.10}, \eqref{6.41} and \eqref{6.50} yield 
$\lim \cal M_2^m=0$. 
Therefore, the pair $(v=v^0,\,\zeta=\zeta_0)$ satisfies the equation 
\eqref{6.6}, and the theorem is proved.
$\blacksquare$

 Theorem 6.1 is a generalization of the result  obtained in \cite{17} for a model of non-linear viscous fluid. 

\section{Coupled problem for the viscosity function $\phi_1$.}
For the viscosity function $\phi_1$ defined by \eqref{1.11}, we take the
equation of the balance of thermal energy in the following form:
\begin{equation}\label{7.1}
\chi\sum_{i=1}^n\frac{\di^2\tau}{\di x_i^2}+2\eps\phi_1(\vert E\vert,\mu(u,E),
I(Pu),\tau)I(Pu)-u_i\frac{\di\tau}{\di x_i}=0 \quad \text{in } \Omega,
\end{equation}
where 
\begin{equation}\label{7.2}
\phi_1(\vert E\vert,\mu(u,E),I(Pu),\tau)=b(\vert E\vert,\mu(u,E),\tau)(\lambda
+I(Pu))^{-\frac12}+\psi(I(Pu),\vert E\vert,\mu(u,E),\tau),
\end{equation}
and $P$ is an operator of regularization determined in \eqref{3.12}, 
\eqref{3.13}.

The presence of the operator $P$ in \eqref{7.1} means that the model is not
local and it is natural from the physical point of view (see Section 3).

The motion equations, the condition of incompressibility and boundary 
conditions are given by \eqref{2.1} under $k=1$, \eqref{2.2}, and 
\eqref{2.4}--\eqref{2.6}.

We assume that \eqref{2.7}--\eqref{2.9} hold and the functions $b$ and $\psi$
satisfy the following conditions which are similar to the conditions (C1)
and (C2).

\begin{description}
\item[(C1a)]
$b:(y_1,y_2,y_3)\to b(y_1,y_2,y_3)$ is a function continuous in $\R_+\times
   [0,1]\times \R$, and in addition
\begin{equation}\label{7.3}
0\le b(y_1,y_2,y_3)\le a_0, \qquad (y_1,y_2,y_3)\in\R_+\times[0,1]\times\R. 
\end{equation}
\item[(C2a)]
$\psi:(y_1,y_2,y_3,y_4)\to\psi(y_1,y_2,y_3,y_4)$ is a function continuous
 in $\R_+\times\R_+\times[0,1]\times\R$, and for an arbitrary fixed $(y_2,
y_3,y_4)\in\R_+\times[0,1]\times\R$ the function $\psi(.,y_2,y_3,y_4):y_1
\to\psi(y_1,y_2,y_3,y_4)$ is continuously differentiable in $\R_+$, and the
following inequalities hold
\begin{align}
&a_2\ge\psi(y_1,y_2,y_3,y_4)\ge a_1,    \label{7.4}\\
&\psi(y_1,y_2,y_3,y_4)+2\frac{\di\psi}{\di y_1}(y_1,y_2,y_3,y_4)y_1\ge a_3,
                                        \label{7.5}\\ 
&\Big\vert\frac{\di\psi}{\di y_1}(y_1,y_2,y_3,y_4)\Big\vert y_1\le a_4
                                        \label{7.6}
\end{align}
\end{description}
We introduce mappings $N_1:X\times H_0^1(\Omega)\to X^*$ and $A_1:X\times 
H_0^1(\Omega)\to H^{-1}(\Omega)$ as follows:
\begin{gather}
(N_1(v,\zeta),h)=2\int_\Omega[b(\vert E\vert,\mu(\tilde u+v,E),\tilde\tau+
        \zeta)(\lambda+I(\tilde u+v))^{-\frac12}     \notag \\
+\psi(I(\tilde u+v),\vert E\vert,\mu(\tilde u+v,E),\tilde\tau+\zeta)]\eps_{ij}   
        (\tilde u+v)\eps_{ij}(h)\,dx,                 \notag \\
v,h\in X, \qquad \zeta\in H_0^1(\Omega),             \label{7.7} \\
(A_1(v,\zeta),\xi)=-\int_\Omega(\tilde u_i+v_i)\frac{\di(\tilde\tau+\zeta)}
        {\di x_i}\,\xi\,dx                            \notag \\
+2\eps\int_\Omega\phi_1(\vert E\vert,\mu(\tilde u+v,E),I(P(\tilde u+v)),\tilde
        \tau+\zeta)I(P(\tilde u+v))\xi\,dx-\chi\pi(\tilde\tau,\xi), \notag \\
v\in X, \qquad \zeta,\xi\in H_0^1(\Omega).                    \label{7.8}
\end{gather}
Consider the following problem: find a triple $(v,p,\zeta)$ such that
\begin{gather}
(v,p,\zeta)\in X\times L_2(\Omega)\times H_0^1(\Omega),  \label{7.9}\\
(N_1(v,\zeta),h)-(B^*\,p,h)=(K+F,h), \quad h\in X,      \label{7.10}\\
(Bv,q)=0, \qquad q\in L_2(\Omega),                      \label{7.11}\\
\pi(\zeta,\xi)-\frac1{\chi}\Big(A_1(v,\zeta),\xi\Big)=0, \quad
              \xi\in H_0^1(\Omega).                     \label{7.12} 
\end{gather}
Here we use the notations \eqref{3.19}.

If $(v,p,\zeta)$ is a solution of the problem \eqref{7.9}--\eqref{7.12},
then $(\tilde u+v,p,\tilde\tau+\zeta)$ is a generalized solution of the
problem \eqref{2.1} for $k=1$, \eqref{2.2}, \eqref{2.4}--\eqref{2.6} and
\eqref{7.1}.

\begin{th}
Let $\Omega$ be a bounded domain in $\R^n$, $n=2$ or $3$, with a boundary
$S$ of the class $C^1$, and suppose the conditions  $(C1a)$, $(C2a)$ and 
\eqref{2.7}, \eqref{2.9} are satisfied.

Assume also that there exists a function $\tilde\tau$ such that \eqref{2.8}
holds and in addition
\begin{equation}\label{7.13}
\tilde\tau\in L_\infty(\Omega).
\end{equation}
Then exists a solution of the problem \eqref{7.9}--\eqref{7.12}
\end{th}

\section{Lemmas and estimation of the term generated by the convection.}
Below we suppose that $\Omega$ is a bounded domain in $\R^n$, $n=2$ or $3$
with a boundary $S$ of the class $C^1$. 
\begin{lem}
Let $\rho(x)$ be the distance from $x$ to $S$. Then for an arbitrary 
$\gamma>0$  there exists a function $\alpha_\gamma\in C^1(\overline\Omega)$
such that
\begin{align}
&\alpha_\gamma=1\quad \text{in some vicinity of } S \text{ depending on }
               \gamma,                               \label{8.1}\\
&\alpha_\gamma(x)=0 \quad \text{at } \rho(x)\ge 2\delta(\gamma), \quad
               \delta(\gamma)=\exp(-\frac1\gamma),   \label{8.2}\\
&\Big\vert\frac{\di}{\di x_k}\,\alpha_\gamma(x)\Big\vert\le\frac{\gamma}
               {\rho(x)} \quad \text{at } \rho(x)\le 2\delta(\gamma), \quad
                k=1,\dots,n.                          \label{8.3}  
\end{align}
\end{lem} 
The function $\alpha_\gamma$ satisfying the above conditions was constructed
by Hopf \cite{4}, see also \cite{7}.
\begin{lem}
There exists a constant $c$ depending only on $\Omega$ such that
\begin{equation}\label{8.4}
\Big\|\frac1\rho\,w\Big\|_{L_2(\Omega)}\le c\|w\|_1, \qquad w\in H_0^1(\Omega).
\end{equation}
\end{lem}
This lemma is proved by a partition of unity and local maps followed by the 
application of the Hardy inequality, see e.g. \cite{18}, Lemma 1.10, Chapter
2.
\begin{lem}
Suppose a function $\tilde\tau$ satisfies \eqref{2.8} and \eqref{7.13}.
Then for an arbitrary $\beta>0$ one can construct a function $G$ such that
\begin{equation}\label{8.5}
G\in H^1(\Omega)\cap L_\infty(\Omega), \qquad G\big\vert_S=\tilde\tau\big
     \vert_S,                           
\end{equation}
and in addition
\begin{equation}\label{8.6}
\Big\|\zeta\frac{\di G}{\di x_i}\Big\|_{L_{\frac65}(\Omega)}\le\beta\|
     \zeta\|_1, \qquad \zeta\in H_0^1(\Omega), \quad i=1,\dots,n.
\end{equation}
\end{lem}
{\bf Proof}. Consider the function
\begin{equation}\label{8.7}
G=\alpha_\gamma\,\tilde\tau.
\end{equation}
This function satisfies \eqref{8.5}. It follows from \eqref{7.13}, \eqref{8.7}
and Lemma 8.1 that
\begin{gather}
\Big\vert\frac{\di G}{\di x_i}(x)\Big\vert=\Big\vert\Big(\frac{\di}{\di x_i}
     (\alpha_\gamma\tilde\tau)\Big)(x)\Big\vert\le\frac{\gamma}{\rho(x)}
      \vert\tilde\tau(x)\vert+c\Big\vert\frac{\di\tilde\tau}{\di x_i}\,
      (x)\Big\vert                      \notag\\
\le c_1\Big(\frac{\gamma}{\rho(x)}+\Big\vert\frac{\di\tilde\tau}{\di x_i}
      (x)\Big\vert\Big) \quad \text{at} \quad \rho(x)\le 2\delta(\gamma), 
      \quad i=1,\dots,n.                     \label{8.8}
\end{gather}
\eqref{8.8} implies
\begin{gather}
\Big\|\zeta\frac{\di G}{\di x_i}\Big\|_{L_{\frac65}(\Omega)}\le c_1\Big(\gamma
    \Big\|\frac{\zeta}{\rho}\Big\|_{L_{\frac65}(\Omega)}+\Big(\int_
    {\rho\le 2\delta(\gamma)}\Big\vert\zeta\frac{\di\tilde\tau}{\di x_i}\Big
    \vert^{\frac65}\,dx\Big)^{\frac56}\Big)        \notag\\
\le c_2\Big(\gamma\Big\|\frac{\zeta}{\rho}\Big\|_{L_2(\Omega)}+\|\zeta\|_
     {L_3(\Omega)}\sigma_i(\gamma)\Big),             \label{8.9}
\end{gather}
where
\begin{equation}\label{8.10}
\sigma_i(\gamma)=\Big(\int_{\rho\le 2\delta(\gamma)}\Big(\frac{\di\tilde\tau}
         {\di x_i}\Big)^2\,dx\Big)^{\frac12}, \qquad i=1,\dots,n.
\end{equation}
Here we have applied the H\"older inequality with the indexes $\frac25$ and
$\frac35$ to the integral on the right-hand side of \eqref{8.9}.

\eqref{8.9} and Lemma 8.2 yield
\begin{equation}\label{8.11}
\Big\|\zeta\frac{\di G}{\di x_i}\Big\|_{L_{\frac65}(\Omega)}\le c_3
       \|\zeta\|_1(\gamma+\sigma_i(\gamma)), \qquad i=1,\dots,n.
\end{equation}
Since $\frac{\di\tilde\tau}{\di x_i}\in L_2(\Omega)$, we obtain from 
\eqref{8.10} that $\sigma_i(\gamma)\to 0$ as $\gamma\to 0$, and the Lemma 
is proved.
$\blacksquare$

We consider a mapping $A_2:V\times H_0^1(\Omega)\to H^{-1}(\Omega)$ that is
defined as follows:
\begin{equation}\label{8.12}
(A_2(v,\zeta),\xi)=-\int_\Omega(\tilde u_i+v_i)\frac{\di(G+\zeta)}{\di x_i} 
\,\xi\,dx, \quad v\in X,\quad \zeta,\xi\in H_0^1(\Omega). 
\end{equation}
\begin{lem}
Suppose the conditions \eqref{2.7}, \eqref{2.8} and \eqref{7.13} are satisfied.
Then for an arbitrary $\eta>0$ one can determine the function $G$ so that
\eqref{8.5} is satisfied, and in addition
\begin{equation}\label{8.13}
\vert(A_2(v,\zeta),\zeta)\vert\le c\|\zeta\|_1+\eta(\|v\|_X^2+\|\zeta\|_1^2),
          \qquad v\in V, \quad \zeta\in H_0^1(\Omega).
\end{equation}
\end{lem}
{\bf Proof}. It follows from \eqref{8.12} that
\begin{equation}\label{8.14}
(A_2(v,\zeta),\zeta)=M_1+M_2+M_3,
\end{equation}
where
\begin{gather}
M_1=-\int_\Omega\tilde u_i\frac{\di G}{\di x_i}\,\zeta\,dx, \qquad
M_2=-\int_\Omega v_i\frac{\di G}{\di x_i}\,\zeta\,dx,  \notag\\
M_3=-\int_\Omega(\tilde u_i+v_i)\frac{\di\zeta}{\di x_i}\,\zeta\,dx.
                                                 \label{8.15}                      
\end{gather}
\eqref{2.7} and \eqref{8.6} yield
\begin{align}
&\vert M_1\vert\le\sum_{i=1}^n\|\tilde u_i\|_{L_6(\Omega)}\Big\|\frac{\di
G}{\di x_i}\,\zeta\Big\|_{L_{\frac65}(\Omega)}\le\beta c_1\|\tilde u\|_
      {H^1(\Omega)^n}\|\zeta\|_1\le c_2\|\zeta\|_1,    \label{8.16}\\
&\vert M_2\vert\le\sum_{i=1}^n\|v_i\|_{L_6(\Omega)}\Big\|\frac{\di G}{\di 
      x_i}\,\zeta\Big\|_{L_{\frac65}(\Omega)}\le\beta c_3\|v\|_X\|\zeta\|_1
      \le\frac12\beta c_3(\|v\|_X^2+\|\zeta\|_1^2).          \label{8.17} 
\end{align}
Since $\diver(\tilde u +v)=0$, we obtain by integration by parts that
\begin{equation}\label{8.18}
M_3=-\int_\Omega(\tilde u_i+v_i)\frac{\di\zeta}{\di x_i}\,\zeta\,dx=\int_
      \Omega(\tilde u_i+v_i)\zeta\frac{\di\zeta}{\di x_i}\,dx=0.
\end{equation}
Now \eqref{8.13} follows from  \eqref{8.16}--\eqref{8.18} and Lemma 8.3.
$\blacksquare$

\section{Proof of Theorem 7.1.}
1) It follows from the proof of Theorem 4.1, that if $(v,p,\zeta)$ is a 
solution of the problem \eqref{7.9}--\eqref{7.12}, then
\begin{equation}\label{9.1}
\|v\|_X\le\frac1{\mu_3}(\|K+F\|_{V^*}+\mu_4).
\end{equation}
By \eqref{3.12}, \eqref{3.13} and \eqref{9.1}, we conclude that there exists 
a constant $b_1$  such that
\begin{equation}\label{9.2}
\|I(P(\tilde u+v))\|_{C(\overline\Omega)}\le b_1.
\end{equation}
Consider a function $\phi_3$ such that $\phi_3$ is a function continuous and
bounded in $\R_+\times[0,1]\times\R_+\times\R\to\R_+$ and in addition

\begin{equation}
\phi_3(z)=
\begin{cases}
&\phi_1(z)\text{ at }z=(z_1,z_2,z_3,z_4)\in\R_+\times[0,1]\times[0,b_1]\times\R,\\
&0 \text{ at }z=(z_1,z_2,z_3,z_4)\in\R_+\times[0,1]\times(b_2,\infty)
                      \times\R,   \label{9.3}
\end{cases}
\end{equation}
where $b_2>b_1$ and $\phi_1$  is the function defined by \eqref{7.2}, i.e.
\begin{equation}\label{9.4}
\phi_1(z)=b(z_1,z_2,z_4)(\lambda+z_3)^{-\frac12}+\psi(z_3,z_1,z_2,z_4).
\end{equation}

Define operators $N_2:X\times H_0^1(\Omega)\to X^*$ and $A_3:X\times H_0^1
(\Omega)\to H^{-1}(\Omega)$ by 
\begin{gather}
(N_2(v,\zeta),h)=2\int_\Omega[b(\vert E\vert,\mu(\tilde u+v,E),G+\zeta)
                 (\lambda+I(\tilde u+v))^{-\frac12}      \notag\\
+\psi(I(\tilde u+v),\vert E\vert,\mu(\tilde u+v,E),G+\zeta)]\eps_{ij}
                 (\tilde u+v)\eps_{ij}(h)\,dx,           \notag\\ 
               v,h\in X,\quad \zeta\in H_0^1(\Omega),     \label{9.6}\\
(A_3(v,\zeta),\xi)=2\eps\int_\Omega\phi_3(\vert E\vert,\mu(\tilde u+v,E),
                I(P(\tilde u+v)),G+\zeta)I(P(\tilde u+v))\xi\,dx-\chi
                \pi(G,\xi),                              \notag\\
v\in X, \quad \zeta,\xi\in H_0^1(\Omega),                \label{9.7}
\end{gather}
where $G$ is the function defined in Lemma 8.3.

We consider the problem: find a triple of functions $(v,p,\theta)$ satisfying
\begin{align}
&(v,p,\theta)\in X\times L_2(\Omega)\times H_0^1(\Omega),    \label{9.8}\\
&(N_2(v,\theta),h)-(B^*\,p,h)=(K+F,h), \qquad h\in X,        \label{9.9}\\
&(Bv,q)=0, \qquad q\in L_2(\Omega),                          \label{9.10}\\
&\pi(\theta,\xi)-\frac1{\chi}[(A_2(v,\theta),\xi)+(A_3(v,\theta),\xi)]=0, 
           \qquad \xi\in H_0^1(\Omega).                      \label{9.11}
\end{align}
It follows from the proof of theorem 4.1 that the inequality \eqref{9.1}
is also fulfilled for the solution of the problem \eqref{9.8}--\eqref{9.11}.
Therefore, \eqref{9.2} holds, and if $(v,p,\theta)$ is a solution of the 
problem \eqref{9.8}--\eqref{9.11}, then the triple $(v,p,\zeta=G-\tilde\tau+
\theta)$ is a solution of the problem \eqref{7.9}--\eqref{7.12}.

The problem \eqref{9.8}--\eqref{9.11} is equivalent to the following problem:
find a pair $(v,\theta)$ such that
\begin{align}
&(v,\theta)\in V\times H_0^1(\Omega),                     \label{9.12}\\
&(N_2(v,\theta),h)=(K+F,h),  \qquad h\in V,               \label{9.13}\\
&\pi(\theta,\xi)-\frac1\chi[(A_2(v,\theta),\xi)+(A_3(v,\theta),\xi)]=0,
                             \qquad \xi\in H_0^1(\Omega). \label{9.14}
\end{align}
Indeed, if $(v,p,\theta)$ is a solution of the problem 
\eqref{9.8}--\eqref{9.11}, then the pair $(v,\theta)$ is a solution of the 
problem \eqref{9.12}--\eqref{9.14}. Conversely, let $(v,\theta)$ be a 
solution of the problem \eqref{9.12}--\eqref{9.14}. By virtue of \eqref{9.13}
and Lemma 4.2 there exists a unique function $p\in L_2(\Omega)$ such that the
triple $(v,p,\theta)$ is a solution of the problem \eqref{9.8}--\eqref{9.11}.

2) Let us prove the existence of a solution of the problem 
\eqref{9.12}--\eqref{9.14}.

We define a mapping $U:V\times H_0^1(\Omega)\to V^*\times H^{-1}(\Omega)$ by
\begin{gather}
(U(v,\theta),(h,\xi))=(N_2(v,\theta),h)+\pi(\theta,\xi)-\frac1\chi[(A_2(v,
     \theta),\xi)+(A_3(v,\theta),\xi)]-(K+F,h),            \notag\\
v,h\in V, \qquad \theta,\xi\in H_0^1(\Omega).               \label{9.15}
\end{gather}
It is easy to verify that the problem \eqref{9.12}--\eqref{9.14} is 
equivalent to the following one: find a pair $(v,\theta)$ such that
\begin{align}
&(v,\theta)\in V\times H_0^1(\Omega),                \label{9.16}\\
&(U(v,\theta),(h,\xi))=0,  \qquad (h,\xi)\in V\times H_0^1(\Omega).
                                                     \label{9.17}
\end{align}
Indeed, taking the pair $(h,0)$ in \eqref{9.17}, we obtain \eqref{9.13},
and taking the pair $(0,\xi)$, we get \eqref{9.14}. Conversely, by adding
\eqref{9.13} and \eqref{9.14}, we get \eqref{9.17}.

Let $\{V_m\}$ and $\{Z_m\}$ be sequences of finite-dimensional subspaces in
$V$ and $H_0^1(\Omega)$, respectively, such that
\begin{align}
&\lim_{m\to\infty}\inf_{g\in V_m}\|u-g\|_X=0, \qquad u\in V, \label{9.18}\\
&\lim_{m\to\infty}\inf_{h\in Z_m}\|w-h\|_1=0, \qquad w\in H_0^1(\Omega),
                                                             \label{9.19}\\
&V_m\subset V_{m+1}, \qquad Z_m\in Z_{m+1}.                   \label{9.20}
\end{align}
We search for an approximate solution $v_m$, $\theta_m$ of the problem
\eqref{9.12}--\eqref{9.14} in the form
\begin{align}
&(v_m,\theta_m)\in V_m\times Z_m,                    \label{9.21}\\
&(N_2(v_m,\theta_m),h)=(K+F,h), \qquad h\in V_m,     \label{9.22}\\
&\pi(\theta_m,\xi)-\frac1\chi[(A_2(v_m,\theta_m),\xi)+(A_3(v_m,\theta_m),
                   \xi)]=0,     \qquad \xi\in Z_m.    \label{9.23}
\end{align}
\eqref{9.3} and \eqref{9.7} imply
\begin{equation}\label{9.24}
\vert(A_3(v,\theta),\theta)\vert\le c_1\int_\Omega\vert\theta\vert\,dx+
      \chi(\pi(G,G))^{\frac12}\|\theta\|_1\le c_2\|\theta\|_1.
\end{equation}
Inequality \eqref{4.3} is also realized for the operator $N_2$. So that by
using \eqref{8.13}, \eqref{9.15} and \eqref{9.24}, we obtain
\begin{gather}
(U(v,\theta),(v,\theta))\ge\mu_3\|v\|_X^2+\|\theta\|_1^2-c_3\|v\|_X-c_4
    \|\theta\|_1-\frac{\eta}{\chi}(\|v\|_X^2+\|\theta\|_1^2),  \notag\\
(v,\theta)\in V\times H_0^1(\Omega).              \label{9.25}
\end{gather}
By Lemma 8.4 we can reckon that
\begin{equation}\notag
\eta=\frac12\,\chi\min(\mu_3,1).
\end{equation}
Then there exists a constants $r>0$ such that
\begin{equation}\label{9.26}
(U(v,\theta),(v,\theta))\ge 0, \quad \text{if}\quad \|v\|_X+\|\theta\|_1\ge r.
\end{equation}
Therefore, for an arbitrary $m\in\N$ there exists a pair $(v_m,\theta_m)$
satisfying
\begin{equation}\label{9.27}
(v_m,\theta_m)\in V_m\times Z_m, \quad (U(v_m,\theta_m),(h,\xi))=0, \quad
                  (h,\xi)\in V_m\times Z_m.
\end{equation}
This pair is the solution of the problem \eqref{9.21}--\eqref{9.23}, 
 and moreover
\begin{equation}\label{9.28}
\|v_m\|_X+\|\theta_m\|_1\le r.
\end{equation}
It follows from the proof of Theorem 4.1 that
\begin{equation}\label{9.29}
\|v_m\|_X\le\frac1{\mu_3}(\|K+F\|_{V^*}+\mu_4).
\end{equation}

3) By virtue of \eqref{9.28}, we can extract a subsequence $\{v_k,\theta_k\}$
from the sequence $\{v_m,\theta_m\}$ such that
\begin{align}
&v_k\rightharpoonup v_0 \quad \text{in}\quad V,            \label{9.30}\\
&v_k\to v_0 \quad \text{in} \quad L_4(\Omega)^n \quad \text{and a.e. in }
                  \quad    \Omega,                          \label{9.31}\\
&\theta_k\rightharpoonup\theta_0 \quad \text{in} \quad H_0^1(\Omega), 
                                                           \label{9.32}\\
&\theta_k\to\theta_0 \quad \text{in} \quad L_2(\Omega)\quad \text{and a.e. 
                in}\quad \Omega,                           \label{9.33}\\
&N_2(v_k,\theta_k)\rightharpoonup\cal F \quad \text{in} \quad X^*.
                                                            \label{9.34}
\end{align}
Let $k_0$ be a fixed positive number, and $h\in V_{k_0}$. Observing 
\eqref{9.34}, we pass to the limit in \eqref{9.22} with $m$ replaced by $k$,
and obtain
\begin{equation}\label{9.35}
(\cal F,h)=(K+F,h), \qquad h\in V_{k_0}.
\end{equation}
Since $k_0$ is an arbitrary positive integer, we get by \eqref{9.18} and
\eqref{9.35} that
\begin{equation}\label{9.36}
\cal F=K+F \quad \text{in} \quad V^*.
\end{equation}
We present the operator $N_2$ in the form
\begin{equation}\label{9.37}
N_2(v,\zeta)=\tilde N(v,v,\zeta),
\end{equation}
where the operator $(v,w,\zeta)\to\tilde N(v,w,\zeta)$ is considered as a 
mapping of $V\times V\times H_0^1(\Omega)$ into $V^*$ according to
\begin{gather}
(\tilde N(v,w,\zeta),h)=2\int_\Omega[b(\vert E\vert,\mu(\tilde u+w,E),G+
    \zeta)(\lambda+I(\tilde u+v))^{-\frac12}              \notag\\
+\psi(I(\tilde u+v),\vert E\vert,\mu(\tilde u+w,E),G+\zeta)]\eps_{ij}
    (\tilde u+v)\eps_{ij}(h)\,dx,                         \notag\\
v,w\in V, \qquad \zeta\in H_0^1(\Omega).                  \label{9.38}
\end{gather}
For an arbitrary fixed $w\in V$ and $\zeta\in H_0^1(\Omega)$ the operator
$\tilde N(.,w,z):v\to\tilde N(v,w,z)$ is monotone (see \eqref{4.2}).
Therefore,
\begin{equation}\label{9.39}
(\tilde N(v_k,v_k,\theta_k)-\tilde N(z,v_k,\theta_k),v_k-z)\ge 0, \quad
                      z\in V, \quad k\in\N.
\end{equation}
\eqref{9.31}, \eqref{9.33} and Lebesgue theorem give
\begin{equation}\label{9.40}
\lim\|\tilde N(z,v_k,\theta_k)-\tilde N(z,v_0,\theta_0\|_{X^*}=0,
             \qquad z\in V.
\end{equation}
By \eqref{9.22}, \eqref{9.30} and \eqref{9.37}, we obtain
\begin{equation}\label{9.41}
(\tilde N(v_k,v_k,\theta_k),v_k)=(K+F,v_k)\to(K+F,v_0),
\end{equation}
and \eqref{9.18} yields
\begin{equation}\label{9.42}
\lim(\tilde N(v_k,v_k,\theta_k),z)=(K+F,z), \qquad z\in V.
\end{equation}
Observing \eqref{9.30}, \eqref{9.40}--\eqref{9.42}, we pass to the limit in 
\eqref{9.39}.

Then we get
\begin{equation}\label{9.43}
(K+F-\tilde N(z,v_0,\theta_0),v_0-z)\ge 0, \qquad z\in V.
\end{equation}
We choose $z=v_0-\gamma h$, $\gamma>0$, $h\in V$, and consider $\gamma\to 0$.
Then, taking into account that $\tilde N(.,v_0,\theta_0):z\to\tilde N(z,v_0,
\theta_0)$ is a Lipschitz continuous mapping of $V$ into $V^*$ (see 
\eqref{4.1}), we obtain
\begin{equation}\label{9.44}
(K+F-\tilde N(v_0,v_0,\theta_0),h)\ge 0.
\end{equation}
This inequality holds for any $h\in V$. Therefore, replacing $h$ by $-h$
shows that equality holds true in \eqref{9.44}. From here by \eqref{9.30}, 
\eqref{9.32} and \eqref{9.37}, we deduce
\begin{equation}\label{9.45}
(v_0,\theta_0)\in V\times H_0^1(\Omega), \quad (N_2(v_0,\theta_0),h)=(K+
             F,h), \qquad h\in V.
\end{equation}
\eqref{8.12}, \eqref{9.7}, \eqref{9.31}--\eqref{9.33} imply
\begin{align}
&\lim(A_2(v_k,\theta_k),\xi)=(A_2(v_0,\theta_0),\xi),   \notag\\ 
&\lim(A_3(v_k,\theta_k),\xi)=(A_3(v_0,\theta_0),\xi), \qquad 
                  \xi\in H_0^1(\Omega).                  \label{9.46}
\end{align}
Bearing in mind \eqref{9.32} and \eqref{9.46}, we pass to the limit in
\eqref{9.23} for fixed $\xi\in Z_{k_0}$. Then by \eqref{9.19}, we deduce
\begin{equation}\label{9.47}
\pi(\theta_0,\xi)-\frac1\chi[(A_2(v_0,\theta_0),\xi)+(A_3(v_0,\theta_0),
                \xi)]=0, \quad \xi\in H_0^1(\Omega).
\end{equation}
It follows from \eqref{9.45} and \eqref{9.47} that the pair $(v=v_0,\,\,
\theta=\theta_0)$ is a solution of the problem \eqref{9.12}--\eqref{9.14}.
Hence, there exists a unique function $p$ such that the triple $(v=v_0,\,\,
p,\,\,\theta=\theta_0)$ is a solution of the problem \eqref{9.8}--\eqref
{9.11}.

\eqref{9.29} and \eqref{9.30} yield
\begin{equation}\label{9.48}
\|v_0\|_X\le\frac1{\mu_3}(\|K+F\|_{V^*}+\mu_4).
\end{equation}
Since \eqref{9.1} implies \eqref{9.2}, we get $\|I(P(\tilde u+v_0))\|_
{C(\overline\Omega)}\le b_1$. From here by \eqref{7.8}, \eqref{9.3} and 
\eqref{9.7}, we obtain that the triple $(v=v_0,\,\,p,\,\,\zeta=G-\tilde\tau
+\theta_0)$ is a solution of the problem \eqref{7.9}--\eqref{7.12}. The
theorem is proved.
$\blacksquare$

\end{document}